\documentclass[10pt, conference, compsocconf]{IEEEtran}

\usepackage[dvipsnames]{xcolor}
\usepackage{listings}
\usepackage{graphicx}
\usepackage{todonotes}
\usepackage{paralist}

\usepackage[hyphens]{url}
\usepackage[hidelinks]{hyperref}
\hypersetup{breaklinks=true}
\usepackage{cite}

\newcommand\YAMLcolonstyle{\color{red}\mdseries}
\newcommand\YAMLkeystyle{\color{black}\bfseries}
\newcommand\YAMLvaluestyle{\color{blue}\mdseries}

\makeatletter

\newcommand\language@yaml{yaml}

\expandafter\expandafter\expandafter\lstdefinelanguage
\expandafter{\language@yaml}
{
  keywords={true,false,null,y,n},
  keywordstyle=\color{darkgray}\ttfamily\bfseries,
  basicstyle=\YAMLkeystyle\small, 
  sensitive=false,
  comment=[l]{\#},
  morecomment=[s]{/*}{*/},
  commentstyle=\color{purple}\ttfamily,
  stringstyle=\YAMLvaluestyle\ttfamily,
  moredelim=[l][\color{orange}]{\&},
  moredelim=[l][\color{magenta}]{*},
  moredelim=**[il][\YAMLcolonstyle{:}\YAMLvaluestyle]{:}, 
  morestring=[b]',
  morestring=[b]",
  literate =    {---}{{\ProcessThreeDashes}}3
                {>}{{\textcolor{red}\textgreater}}1     
                {|}{{\textcolor{red}\textbar}}1 
                {\ -\ }{{\mdseries\ -\ }}3,
}

\lst@AddToHook{EveryLine}{\ifx\lst@language\language@yaml\YAMLkeystyle\fi}
\makeatother

\usepackage{xspace}

\newcommand\ProcessThreeDashes{\llap{\color{cyan}\mdseries-{-}-}}
\newcommand{\itwokit}{\emph{i2kit}\xspace}

\begin{document}
\title{i2kit: A Tool for Immutable Infrastructure Deployments based on Lightweight Virtual Machines specialized to run Containers}

\author{\IEEEauthorblockN{Pablo Chico de Guzm\'an}
\IEEEauthorblockA{IMDEA Software Institute\\
Madrid, Spain\\
pablo.chico@imdea.org}
\and
\IEEEauthorblockN{Felipe Gorostiaga}
\IEEEauthorblockA{IMDEA Software Institute\\
Madrid, Spain\\
felipe.gorostiaga@imdea.org}
\and
\IEEEauthorblockN{C\'esar S\'anchez}
\IEEEauthorblockA{IMDEA Software Institute\\
Madrid, Spain\\
cesar.sanchez@imdea.org}
}

\maketitle

\begin{abstract}
  Container technologies, like Docker, are becoming increasingly
  popular.
  Containers provide exceptional developer experience because
  containers offer lightweight isolation and ease of software
  distribution.
  Containers are also widely used in production environments, where a
  different set of challenges arise such as security, networking, service
  discovery and load balancing.
  Container cluster management tools, such as Kubernetes, attempt to
  solve these problems by introducing a new control layer with the
  container as the unit of deployment.
  However, adding a new control layer is an extra configuration step
  and an additional potential source of runtime errors.
  The virtual machine technology offered by cloud providers is more
  mature and proven in terms of security, networking, service
  discovery and load balancing.
  However, virtual machines are heavier than containers for local
  development, are less flexible for resource allocation, and suffer
  longer boot times.

  This paper presents an alternative to containers that enjoy the best
  features of both approaches: (1) the use of mature, proven cloud
  vendor technology; (2) no need for a new control layer; and (3) as
  lightweight as containers.
  Our solution is \itwokit, a deployment tool based on the immutable
  infrastructure pattern, where the virtual machine is the unit of
  deployment.
  The \itwokit tool accepts a simplified format of Kubernetes
  Deployment Manifests in order to reuse Kubernetes' most successful
  principles, but it creates a lightweight virtual machine for each
  Pod using Linuxkit.
  Linuxkit alleviates the drawback in size that using virtual machines
  would otherwise entail, because the footprint of Linuxkit is
  approximately 60MB.
  Finally, the attack surface of the system is reduced since Linuxkit
  only installs the minimum set of OS dependencies to run containers,
  and different Pods are isolated by hypervisor technology.
\end{abstract}


\section{Introduction}
\label{sec:intro}

Docker containers~\cite{Merkel-Docker-2014} have popularized the use
of lightweight virtualization technologies such as
LXC~\cite{www-infoworld}.
Some large companies report running all of their services in
containers (e.g.~\cite{clark-everything-in-container}), and Container
as a Service (CaaS) products are available from the main cloud players
including Amazon EC2 Container Service, Azure Container Service, and
Google Container Engine Service.

There are good reasons for the popularity of containers: containers
provide extremely fast instantiation times, small per-instance memory
footprints, high density on a single host and ease of software
distribution.
These features allow containers to provide an exceptional developer
experience.
Developers are able to run third party dependencies such as databases,
message brokers, proxies,\ldots each in its own container.
Additionally, everything is easily integrated with the application
under development with enough isolation and density of containers to
run many small services in the developer's local machine.
In fact, containers have popularized the so-called micro-service
architectures~\cite{fowler-microservices,thones15microservices}.

However, containers also suffer some drawbacks.
Although the high density is of great value in a \emph{local
  environment} or for continuous integration (CI) jobs, containers
introduce new challenges in \emph{production environments} including
security, networking, load balancing and service discovery.
These challenges have already been addressed by traditional cloud
vendor technology using virtual machines (VMs) as the unit of
deployment, but these solutions are not immediately applicable when
the unit of deployment is the container.
Container cluster management tools---like Kubernetes~\cite{Burns:k8},
Docker Swarm~\cite{docker-swarm} and
Mesos~\cite{Hindman:mesos}---attempt to address these issues directly.
Kubernetes combines the goodness of Docker with Google best practices
for massive deployments.
For example, applications are defined using a declarative model based
on Manifest Files, which are becoming the \emph{de facto} standard to
define application behavior at deployment time.
Kubernetes also introduces the notion of a Pod, a group of strongly
related containers that must be deployed on a single host.
Finally, Kubernetes provides service discovery and load balancing in
the context of containers, at the price of a full additional control
plane layer, which requires an additional setup step and a sensitive
runtime dependency that runs in the user's infrastructure.
Kubernetes also imposes a non-negligible learning curve, and debugging
its control plane when things go wrong can become extremely hard.

We argue in this paper that in production environments there is a
better alternative than high density of containers.
The first reason is that container isolation is weaker than the
isolation provided by hypervisors, which has been reported as a
security concern~\cite{docker-security}.
Second, although containers are very flexible in order to optimize the
use of infrastructure resources, modern VM technologies offer machines
with increasingly small sizes (even a few hundred Mb) and that can be
booted by seconds.
The reduced size of Linux distributions specialized for running
containers (Linuxkit is able to create Linux distributions with a 60MB
footprint) reduces this difference in size and booting time, which
diminishes the advantages of using high density of containers in
production environments.
  
This paper presents \itwokit, an open source tool for immutable
infrastructure deployments.
The \itwokit tool reads a simplified version of Kubernetes Deployment
Manifest Files, and transforms each Pod into a specialized virtual
machine using Linuxkit.
The \itwokit tool applies AWS technology such as Elastic Load
Balancers, Auto Scalability Groups and Route 53 Domains to solve the
problems of networking, service discovery and load balancing (other
cloud vendors technology could be easily supported).
The key idea behind \itwokit is to keep all the good features of
Docker for local development and CI, and remove the challenges that
container high density introduce in production environments.
    
The rest of the paper is organized as follows:
Section~\ref{sec:k8} briefly describes Kubernetes and analyzes some of
the challenges when adopting the Kubernetes technology.
Section~\ref{sec:i2kit-high-level} introduces \itwokit and explains
how to map Kubernetes Deployments into native cloud vendor resources.
Section~\ref{sec:i2kit-implementation} describes the \itwokit
implementation by following a Kubernetes Deployment example.
Section~\ref{sec:exp} measures the impact of \itwokit on different
metrics such as security, booting times, network-performance and
cluster memory consumption.
Finally, Section~\ref{sec:conclusions} concludes and describes some
research lines for future work.

\section{Kubernetes. Advantages and Disadvantages}
\label{sec:k8}

Kubernetes is an evolution of the Borg~\cite{borg} and
Omega~\cite{omega} cluster manager tools, adapted for containers.
We first introduce the best features of Kubernetes, which we adapt to
\itwokit, and then we analyze some weak aspects of Kubernetes, which
\itwokit is designed to target.

\subsection{Kubernetes Desirable Features}
\label{sec:k8-good}

\begin{figure}[t]
    \centering\includegraphics[scale=0.8]{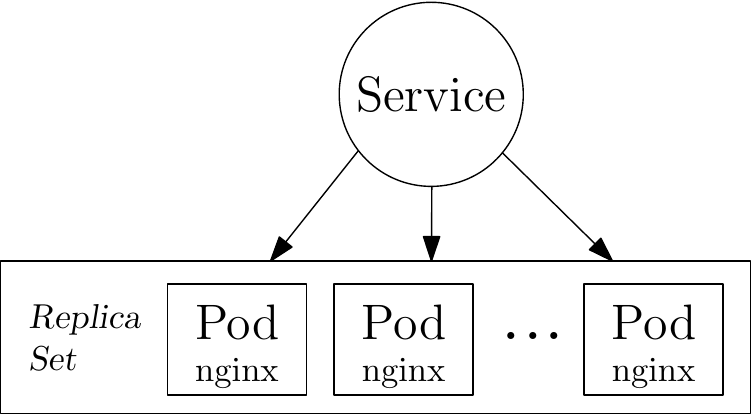}
  \caption{High-Level view of a Kubernetes Deployment.}
  \label{fig:k8-deployment}
  \end{figure}

Kubernetes deploys applications using three main entities: Pods,
Replica Sets and Services, which are defined in Manifest Files
following a declarative model.
Fig.~\ref{fig:k8-deployment} shows the relations between these
entities for the Kubernetes Deployment Manifest shown in
Fig.~\ref{fig:nginx-kubernetes}.
The next subsections describe in detail these concepts and how they
are related.

   \begin{figure}[t]
 %  \begin{lstlisting}[language=yaml,basicstyle=\small]
   \begin{lstlisting}[language=yaml]
 apiVersion: apps/v1beta2
 kind: Deployment
 metadata: { name: nginx-deployment }
 spec:
   selector:
     matchLabels: { app: nginx }
   replicas: 3
   template:
     metadata: { labels: { app: nginx} }
     spec:
       containers: 
       - name: nginx
         image: nginx:1.7.9
         ports: [ containerPort: 80 ]
   \end{lstlisting}
   \caption{\emph{Nginx} Manifest File.}
   \label{fig:nginx-kubernetes}
   \end{figure}

%   \begin{figure}[t]
% %  \begin{lstlisting}[language=yaml,basicstyle=\small]
%   \begin{lstlisting}[language=yaml]
% apiVersion: apps/v1beta2
% kind: Deployment
% metadata:
%   name: nginx-deployment
% spec:
%   selector:
%     matchLabels:
%       app: nginx
%   replicas: 3
%   template:
%     metadata:
%       labels:
%         app: nginx
%     spec:
%       containers:
%       - name: nginx
%         image: nginx:1.7.9
%         ports:
%         - containerPort: 80
%   \end{lstlisting}
%   \caption{\emph{Nginx} Manifest File.}
%   \label{fig:nginx-kubernetes}
%   \end{figure}
  
\subsubsection{Declarative Model}
\label{sec:declarative-model}

The declarative model of Deployment Manifest Files works as follows:
  \begin{itemize}
  \item The user declares the desired state of his application in a
    YAML Manifest File.
    Manifest Files include a description of which container images to
    run, the ports exposed by each container, the number of replicas
    and policies defining how to perform rolling updates.
  \item The Kubernetes Control Plane receives Manifest Files via its
    API, which are then recorded as part of the cluster's combined
    desired state.
  \item The Kubernetes Control Plane issues workloads to the nodes in
    the cluster.
    In our example, the Kubernetes control plane creates three
    \emph{nginx} Pods, a Replica Set to ensure that three instances of
    the \emph{nginx} Pod are always running, and a Service to load
    balance incoming traffic between these Pods.
  \item The Kubernetes Control Plane watches the Pod state in order to
    restore the desired state in case of a failure.
  \end{itemize}

There is a significant difference between the declarative approach
described above and an imperative language to describe control planes.
In an imperative model the user issues a procedure with specific
commands to reach the desired state.
A declarative description is usually much shorter and simpler than a
long sequence of imperative commands, and describes the details of how
to create and coordinate the different Kubernetes objects.
  
Occasionally, the cluster might evolve into an undesired state.
Since Kubernetes keeps the desired state, it can then take actions to
reconcile the current state and the desired state of the cluster.
For example, if a node in the cluster becomes unreachable, the
Kubernetes Control Plane will reschedule the Pods running on the node
in a healthy node, reconciling the current and the desired state.

\subsubsection{Pods}
\label{sec:pods}

The unit of deployment in Kubernetes is not the container, but the Pod
of containers. 
There are advanced use-cases that justify the run of multiple
containers inside a single Pod.
For example:
\begin{itemize}
\item a log scraper tailing the output of the user container to a
  centralized logging service.
\item a stats collector sending metrics to perform analytics.
\item a sidecar container providing features for the user container.
\end{itemize}

The Pod is essentially a sandbox that allows running containers inside.
Informally, a Pod ring-fences an area of the host OS, builds a network
stack, creates a bunch of kernel name-spaces, and runs one or more
containers.
Containers running in the same Pod share the same environment.
For example, all containers in the same Pod share the same IP address,
so if the containers need to communicate they can simply use the Pod
localhost interface.

The deployment of a Pod is an all-or-nothing job.
A Pod is never declared as up and available until every container is
up and running.
A Pod can only exist on a single node, even if the Pod runs multiple
containers.

Pods is in some way an abstraction for having a dedicated machine for
the containers in the Pod, without dealing with the high density of
containers running on the same host.  
%
%\todo[inline]{Cesar: I am not sure I understand this point about dealing with
%  the high density}
%\todo[inline]{Pablo: things like port collision, access via localhost...}

\subsubsection{Replica Sets}
\label{sec:replica-sets}

A Replica Set builds on top of a set of Pods. 
A Replica Set takes a Pod template and instantiates the desired number
of replicas of the Pod.
Replica Sets instantiate a background reconciliation loop that ensures
that the right number of replicas are always running, forcing the
reconciliation between the desired state and the current state.
In this sense, Kubernetes applies the immutable infrastructure
principle~\cite{i2} at the Pod level instead of at the host level.

\subsubsection{Services}
\label{sec:services}

Pods are mortal and, in practice it is not unusual that a given Pod
dies.
On failure, the dying Pod is replaced by a new Pod, which probably
runs on a different node with a different IP.
Moreover, when performing rolling updates it is common that the new
Pods have different IPs than the old Pods.
Therefore, the application logic cannot rely on Pod IPs.

The solution for this problem is the use of services which provide a
reliable networking endpoint for a set of Pods.
Services can also load balance the traffic between a set of Pods.

\subsubsection{Deployments}
\label{sec:deployments}

A very common use case of Kubernetes is an application that needs to
run a set of Pods, ensuring a number of running replicas via a Replica
Set, and preforming load balance of the incoming traffic to these
replicas via a Service object.
To facilitate this scenario, Kubernetes provides a higher-level
object called Deployment.
A Deployment is built using Pods, Replica Sets, and Services, making
it transparent the use of these three objects.
Deployments also provides versioning of Manifest Files, rolling updates
strategies and rollback policies.

\subsection{The Drawbacks of Kubernetes}
\label{sec:k8-bad}

\begin{figure}[t]
%    \centering\includegraphics[width=0.95\linewidth]{./Figs/k8-cluster}
    \centering\includegraphics[scale=0.8]{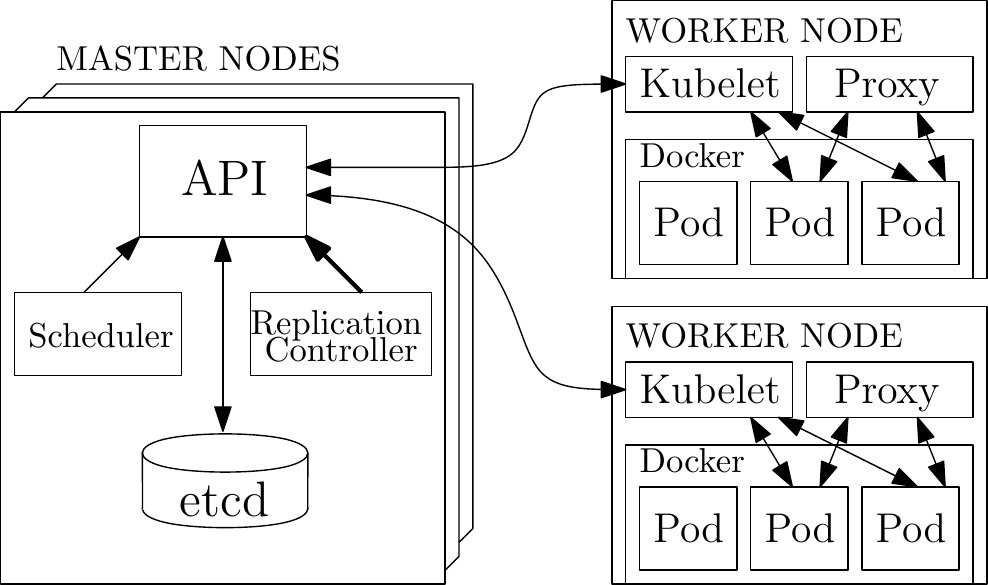}
  \caption{Kubernetes cluster high-level view.}
  \label{fig:k8-cluster}
\end{figure}

Fig.~\ref{fig:k8-deployment} shows a simple Kubernetes Deployment,
built using the concepts described above.
In order to support all these concepts, Kubernetes needs to implement
a complex system.
Fig.~\ref{fig:k8-cluster} gives a more precise view of the
components which form the Kubernetes Control Plane.
  
First, Kubernetes requires a Distributed and Reliable Store Cluster.
The most common solution to this end is \emph{etcd}~\cite{etcd}, a
Key-Value Store based on the Raft~\cite{raft} protocol.
Kubernetes also requires a cluster of Master Nodes.
Master Nodes execute three different components: Api, Scheduler and
Replication Controller.
Also, every Worker Node requires the Kubelet (responsible of executing
the tasks assigned by the Scheduler) and the Kube-Proxy (responsible
of service discovery and load balancing in a high density container
environment).
This complex setup enables powerful deployment scenarios, and it is
arguably a great fit for companies running their own infrastructure
because these companies already need to handle cluster complexity.
However, for the majority of companies and users running on the cloud,
managing such a complex setup is complicated and error prone.
Even further, users need to take into account that the components in
the Kubernetes Control Plane are a runtime dependency for the
applications.
An error in the Kubernetes Control Plane is not only difficult to
debug, but it also disturbs running applications by affecting, for
example, service discovery.

All this complexity has been already solved in the context of virtual
machines and thoroughly tested in practice.
In some sense, Kubernetes is re-implementing similar pre-existing
solutions but for environments with high density of containers per
host.
The consequences of supporting high density of containers are:

\begin{itemize}
\item an additional Control Plane layer which consumes resources and
  adds complexity.
\item it requires virtual networking to provide a unique Pod IP,
  routable from any of the Worker Nodes.
\item it imposes a semi-immutable infrastructure approach.
\end{itemize}
  
Immutable infrastructure~\cite{i2} (\emph{i2}) is an approach to
managing software deployments wherein the servers (where components
run) are replaced rather than changed or modified in every software
update.
Kubernetes reuses Worker Nodes between Pods replacement so it is not
considered to follow the immutable infrastructure principle.
Kubernetes only applies immutable infrastructure principles at the Pod
level, which is sometimes considered insufficient.
For example, Worker Nodes might leak memory and become unreachable
after a number of Pods re-deployments.
Also, it is very common that Worker Nodes become unhealthy due to the
lack of storage resources, for example by the garbage accumulated by
old docker images from previous deployments.
  
\section{I2kit Design}
\label{sec:i2kit-high-level}

The \itwokit  tool aims to enjoy the good features of Docker and
Kubernetes, while removing the security concerns of weaker container
isolation and the extra complexity of self managing a Kubernetes cluster.
In a nutshell:

\begin{itemize}
\item \itwokit uses VMs as the unit of deployment, for
  security and infrastructure maturity reasons;
\item every VM will host a Pod instance;
\item the rest of the Kubernetes objects (Replica Sets, Services,
  Deployments) are replaced by the corresponding native cloud vendor
  technology.
\end{itemize}
Note that \itwokit runs Pods, formed by the same Docker images that
developers run in their local environments.
In this manner, \itwokit preserves the developer workflow untouched.

\begin{figure}[t]
%  \centering\includegraphics[width=0.75\linewidth]{./Figs/aws}
  \centering\includegraphics[scale=0.8]{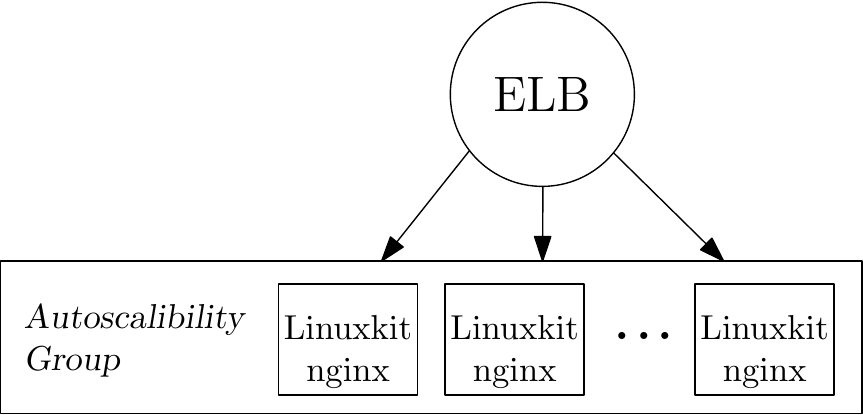}
  \caption{I2kit representation of a Kubernetes Deployment.}
  \label{fig:aws-deployment}
\end{figure}

Fig.~\ref{fig:aws-deployment} shows a high-level view of the Amazon
Web Services created by \itwokit from the Kubernetes Deployment in
Fig.~\ref{fig:k8-deployment}.
The next subsections explain how this transformation is performed.
  
\subsubsection{Mapping Declarative Model}
\label{sec:mapping-declarative-model}

The declarative nature of Kubernetes Manifest Files is a simple and
clean way of defining a deployment behaviour.
The \itwokit tool uses a simplification of Kubernetes Manifest Files
in order to keep its nice declarative features.

\begin{figure}[t]
%  \begin{lstlisting}[language=yaml,basicstyle=\small]
  \begin{lstlisting}[language=yaml]
    name: nginx-deployment
    replicas: 3
    containers:
       nginx:
          { image: nginx:1.7.9 , ports: [ 80 ] }
  \end{lstlisting}
  \caption{\emph{Nginx} \itwokit Manifest File.}
  \label{fig:nginx-i2kit}
  \end{figure}

Fig.~\ref{fig:nginx-i2kit} shows the \itwokit Manifest file
equivalent to the Kubernetes Deployment Manifest shown in
Fig.~\ref{fig:nginx-kubernetes}, which is just a more concise
representation of the same Kubernetes concepts.

The \itwokit implementation is greatly simplified by relying on the
AWS Cloud Formation Service~\cite{aws-cf}.
Cloud Formation receives JSON manifest files to create and manage a
collection of related AWS resources, provisioning and updating them in
an ordered and predictable fashion.
Cloud Formation also follows a declarative description approach.
Cloud Formation templates can specify rolling updates policies to be
applied when the template is modified, allowing the simulation of
Kubernetes rolling updates.
Finally, Cloud Formation keeps a record of the different template
versions, making this information available to \itwokit.
Section~\ref{sec:i2kit-implementation} shows a Cloud Formation
template example generated for the \emph{nginx} Deployment. 
  
\subsubsection{Mapping Pods}
\label{sec:mapping-pods}

Pods are a very useful abstraction of a virtual machine in a node with
high density of containers.
There is a trivial way to map Pods into \itwokit, because every Pod
instance will run in its own VM.
The drawback of this simple mapping is a loss of performance, because
a virtual machine imposes a non-negligible overhead on infrastructure
resources.
However, there are tools to create minimal Linux distributions
specifically crafted to run containers.
The footprint of these distributions can get as small as 60MB
nowadays, a size comparable to container technology.
The tool \itwokit is built on the assumption that the overhead of
running a Pod per virtual machine is acceptable, and this comparison
will keep improving as leaner Linux distributions are developed.
Section~\ref{sec:memory-overhead} discusses this issue.
For example, Unikernels~\cite{Madhavapeddy:Unikernels} have the
promise of reducing this overhead to the minimum by specializing a
Linux kernel for the software that will be running on top.
Note also that a small overhead is even more acceptable in the context of
production environments, where IT operators tend to be generous
allocating resources and where over-provisioning is common.

Additionally, security is an important concern~\cite{docker-security}
when running Pods from different applications in the same node.
The \itwokit approach alleviates these concerns using directly proven
cloud technology built on virtual machines.

The \itwokit tool uses Linuxkit~\cite{linuxkit}, a toolkit for
building custom minimal, immutable Linux distributions. 
Linuxkit reads YAML templates that describe how to build a Linux
distribution. 
Linuxkit templates support the ability to define \emph{services},
which are a set of containers to be run when the VM boots.
Therefore, every attribute of a Pod has a counterpart in
a Linuxkit template. 
Section~\ref{sec:i2kit-implementation} explains this equivalence using
an example.
  
Linuxkit templates also include sections for the \emph{kernel}
filesystem and the \emph{init} processes, making it possible to
install the minimum set of OS dependencies to allow the execution of
containers.
This minimalist approach also improves the security of the system by
reducing its attacking surface, since unnecessary dependencies are
removed.

As a consequence of running every Pod instance in its own specialized
VM, \itwokit has also immediately solved the problem of virtual
networking, since the Pod unique IP is mapped to the node unique IP.
  
\subsubsection{Mapping Replica Sets}
\label{sec:mapping-replica-sets}

A Replica Set ensures that a fix number of Pod instances are always
running.
Replica Sets also replace Pods that get unreachable.
Non surprisingly, in the realm of virtual machines there are also
solutions that perfectly map this behaviour under the assumption of
running a single Pod per VM.
For example, Amazon Web Services offers Amazon Auto Scalability
Groups~\cite{aws-asg}.
Auto Scalability Groups help to maintain the health and availability
of a fleet of Amazon virtual machines, ensuring that the desired
number of VMs is always running. 
If a VM becomes unhealthy, it also gets replaced.

%As explained above, \itwokit creates an Amazon Machine Image using
%Linuxkit.
%%
%Next, an Auto Scalability Group is created that creates virtual
%machines using the previously created Amazon Machine Image.
%%
%The desired number of virtual machines is set to the number of
%replicas specified in the Deployment Manifest.
  
Note that in the event of a rolling update, a new Amazon Machine Image
(AMI) is generated, and the Auto Scalability Group replaces every
existing virtual machine by new VMs running the last AMI built.
The tool \itwokit does follow a pure immutable infrastructure approach
by design.
  
\subsubsection{Mapping Services}
\label{sec:mapping-services}

A Kubernetes Service provides two different functions: (1) it proxies
incoming requests between a set of Pods, and (2) it provides a custom
immutable endpoint which resolves to the associated Pods.
Again, there is an immediate solution in Amazon Web Services for
proving proxy capabilities, by simply using Amazon Load
Balancers~\cite{aws-elb}.

It is possible to automatically attach every virtual machine created
by an Auto Scalability Group to an AWS Load Balancer at creation time.
The AWS Load Balancer provides a reliable endpoint for our set of Pod
instances.
The port configuration of the AWS Load Balancer is created based on
the information contained in the Deployment Manifest.

There is a drawback with this solution, though: AWS Load Balancer
endpoints are not customizable, while the Kubernetes Service endpoints
are.
To overcome this obstacle, \itwokit creates a Route 53 Domain CNAME
entry resolving to the AWS Load Balancer endpoint using the Deployment
Manifest \emph{name} field. 
In this manner, \itwokit provides the same service discovery
mechanism---based on names---than Kubernetes.

%\subsubsection{Mapping Deployments}
%\label{sec:mapping-deployments}  
%
%The tool \itwokit transforms Deployment Manifest into AWS Cloud
%Formation templates.
%%
%First, \itwokit generates a Pod specialized Linux distribution and
%uploads it as an Amazon Machine Image.
%%
%This Amazon Machine Image is used in the Cloud Formation template to
%configure an Auto Scalability Group matching the replicas specified in
%the Kubernetes Deployment Manifest File.
%%
%The Cloud Formation template also defines an AWS Load Balancer
%matching the Pod port configuration and associates it to the previous
%Auto Scalability Group.
%%
%Finally, the Cloud Formation template defines a Route 53 Domain to
%provide a custom reliable endpoint to the Elastic Load balancer, using
%the Kubernetes Deployment name. In this sense, \itwokit is just a tool
%to transform Deployment Manifest into Cloud Formation templates.

\section{I2kit Implementation}
\label{sec:i2kit-implementation}

The \itwokit tool is open source, and it is actively under development
at the IMDEA Software Institute\footnote{\itwokit is available at 
\texttt{www.github.com/pchico83/i2kit}.}.
Currently \itwokit transforms Deployment Manifests into AWS Cloud
Formation templates, but support for other cloud vendors is easy to
implement.
This section describes how \itwokit processes the Deployment Manifest
shown in Fig.~\ref{fig:nginx-i2kit} and creates a Cloud Formation
template.
\begin{figure}[t]
%  \begin{lstlisting}[language=yaml,basicstyle=\small]
  \begin{lstlisting}[language=yaml]
kernel:
  image: linuxkit/kernel:4.9.63
  cmdline: "console=tty0"
init: 
  - linuxkit/init  
  - linuxkit/runc 
  - linuxkit/containerd 
  - linuxkit/ca-certificates
onboot:
  - { name: sysctl, image: linuxkit/sysctl }
  - { name: rngd1,  image: linuxkit/rngd, 
      command: ["/sbin/rngd", "-1"] }
  - { name: dhcpcd, image: linuxkit/dhcpcd }
  - { name: metadata, image: linuxkit/metadata }
services:
  - { name: getty, image: linuxkit/getty, 
      env: [ INSECURE=true] }
  - { name: sshd,  image: linuxkit/sshd }
  - { name: nginx, image: nginx:alpine, 
      capabilities: [ all ] }
trust:
  org: [ linuxkit, library ]
  \end{lstlisting}
  \caption{\emph{Nginx} Linuxkit template.}
  \label{fig:nginx-linuxkit}
\end{figure}
The first step transforms the Pod information contained in a
Deployment Manifest into a Linuxkit template, in order to generate a
minimal Linux distribution specialized in running the Pod containers.
The result is shown in Fig.~\ref{fig:nginx-linuxkit}.
From every container in the Pod, \itwokit extracts the container
relevant information (such as container image, run command,
environment variables) and adds an entry in the \texttt{services}
section of the Linuxkit template.
In our example, this information is:

%\begin{lstlisting}[language=yaml,basicstyle=\small]
\begin{lstlisting}[language=yaml]
services:
   { image: nginx:alpine, capabilities: [all] }
\end{lstlisting}

Note that the value \texttt{all} is used for the capabilities of the
user containers, which is a limitation of the current \itwokit
implementation.
Future work includes equipping \itwokit with an analysis that limits
the capabilities associated to every container.

The remaining fields in the Linuxkit template are pre-generated and
are identical for all Pods.
%, and they are all shared no matter the Pod under deployment.
%
The filesystem of every custom distribution is currently initialized
from the docker image \emph{linuxkit/kernel:4.9.63}.
Also, every custom distribution installs the \emph{init} process,
\emph{runc} and \emph{containerd} to be able to run containers, and
\emph{ca-certificates} to be able to manage certificates.
At boot-time, the containers are executed in sequence order:
\emph{sysctl}, \emph{rngb}, \emph{dhcpcd} and \emph{metadata}.
These are basic services required by any software application. 
Note that \emph{metadata} is installed to be able to manage Amazon
Metadata from the VM itself. 
Then, the containers in the \texttt{services} section run as daemons
in parallel, in particular \emph{getty}, \emph{sshd} and the Pod
containers.
Finally, \itwokit uses content-trust-delivery for images coming from
the \emph{linuxkit} and the \emph{library} organizations.

Once the Linuxkit template has been generated, \itwokit builds the
minimal Linux distribution and uploads it as an Amazon Machine Image.
Assume the id of this AMI is \emph{ami-XXXXX}.
The next step is the generation of the Cloud Formation template, shown
in Fig.~\ref{fig:cloud-formation}.

  \begin{figure}[t]
%  \begin{lstlisting}[language=yaml, basicstyle=\small]
  \begin{lstlisting}[language=yaml]
AWSTemplateFormatVersion: 2010-09-09
Resources:
  LaunchConfig:
    Type: AWS::AutoScaling::LaunchConfiguration
    Properties: { ImageId: ami-XXXXX }
  ASG:
    Type: AWS::AutoScaling::AutoScalingGroup
    Properties:
      LaunchConfigurationName: 
        Ref: LaunchConfig
      MaxSize: 3
      MinSize: 3
      LoadBalancerNames: { Ref: ELB }
  ELB:
    Type: AWS::ElasticLoadBalancing::LoadBalancer
    Properties:
      LoadBalancerName: nginx-deployment
      Listeners:
        LoadBalancerPort: 80
        InstancePort: 80
        Protocol: HTTP
  DNSRecord:
    Type: AWS::Route53::RecordSet
    Properties:
      HostedZoneName: i2kit.com
      Name: nginx-deployment.i2kit.com
      ResourceRecords: 
      -  Fn::GetAtt:("ELB", "DNSName")
      Type: CNAME
  \end{lstlisting}
  \caption{Cloud Formation template for the \emph{nginx} Deployment.}
  \label{fig:cloud-formation}
  \end{figure}

The Cloud Formation template will create four different resources:
\texttt{LaunchConfig}, \texttt{ASG}, \texttt{ELB} and
\texttt{DNSRecord}.
The resource \texttt{LaunchConfig} defines how virtual machines will
be created.
Each machine will use the previously created AMI.
The next resource is \texttt{ASG}, an Auto Scalability Group which use
the \texttt{LaunchConfigurationName} created above in order to create
new VMs.
The minimum and the maximum number of virtual machines matches the
number of replicas in the Deployment Manifest.
Every VM generated by the Auto Scalability Group is associated to an
Elastic Load Balancer defined also in the Cloud Formation template.
\texttt{ELB} stands for the Elastic Load Balancer that takes the name
from the Deployment Manifest \texttt{name} field.
The \texttt{Listeners} information matches the \texttt{ports} section
of the Deployment Manifest, where the \texttt{Protocol} is inferred
from the port number.
Finally, the \texttt{DNSRecord} resource is a CNAME entry for the
Route 53 Domain \emph{i2kit.com}.
This domain is received as a parameter of the \itwokit tool.
The CNAME entry is created based on the Deployment Manifest
\texttt{name} field.
It resolves to the \texttt{ELB} endpoint, providing service discovery
for other deployments.
  
\section{Empirical Evaluation}
\label{sec:exp}

This section compares \itwokit versus the native Kubernetes
implementation based on three different metrics: memory consumption,
network performance, resource fragmentation and booting times.
Additionally, we compare both approaches qualitatively in terms of
security.

\subsubsection{Memory Consumption}
\label{sec:memory-overhead}

\newcommand{\CE}[1]{\multicolumn{1}{c|}{#1}}
\begin{table}[t]
\begin{tabular}{l}
  \begin{tabular}{|l|r|r|r|r|r|} 
    \hline
     Pods & \CE{1}  & \CE{10}      & \CE{20}   & \CE{30} & \CE{40}
    \\\hline\hline                                
    \itwokit     & 78 MB & 0.78 GB  & 1.56 GB & 2.34 GB & 3.12 GB
    \\\hline                                                             
    K8      & 1.94 GB  & 2.09 GB  & 2.27 GB & 2.44 GB  & 2.62 GB
    \\\hline                                                             
\end{tabular} \\
\hspace{8em}(a) Memory comparison \\
  \begin{tabular}{|l|r|r|r|r|r|} 
    \hline
     Pods &   \CE{1}      & \CE{5}   & \CE{25}
    \\\hline\hline                                
    \itwokit     & 129.86 Mbps & 128.191 Mbps  & 128.58 Mbps
    \\\hline                                                             
    K8-1      & 129.17 Mbps  & 25.92 Mbps   & \CE{-}
    \\\hline                                                             
    K8-5      & 108.44 Mbps & 108.36  Mbps & 21.73 Mbps
    \\\hline                                                             
    K8-25      & 97.95 Mbps  & 98.11 Mbps & 97.84 Mbps
    \\\hline                                                             
\end{tabular} \\
\hspace{8em}(b) Network comparison
\end{tabular}
\caption{Comparison of \itwokit vs Kubernetes.}
\label{tab:comparison}
\end{table}

% \begin{table}[t]
%   \centering
%   \begin{tabular}{|l|c|c|c|c|c|} 
%     \hline
%      Pods & 1  & 10      & 20   & 30 & 40
%     \\\hline\hline                                
%     \itwokit     & 78 MB & 0.78 GB  & 1.56 GB & 2.34 GB & 3.12 GB
%     \\\hline                                                             
%     K8      & 1.94 GB  & 2.09 GB  & 2.27 GB & 2.44 GB  & 2.62 GB
%     \\\hline                                                             
% \end{tabular}
% \vspace{5mm}
% \caption{Memory comparing of \itwokit vs. Kubernetes.}
% \label{tab:comparison-memory}
% \end{table}

The overhead that \itwokit imposes for every Pod creation is a
consequence of the overhead of the VM running the Linuxkit
distribution.
In contrast, the Kubernetes overhead for running a Pod is the overhead
of running a Worker Node, which can be shared by several Pods.
Table~\ref{tab:comparison}(a) shows the memory consumption for
running the \emph{nginx} deployment example using different numbers of
Pods replicas.
Table~\ref{tab:comparison}(a) displays the memory footprint of the
total amount of virtual machines that are created (for \itwokit) and
the memory footprint of a single Worker Node running all the Pod
replicas (for Kubernetes).
The Worker Node memory overhead per Pod gets better when more Pods run
on the same Worker Node.

Table~\ref{tab:comparison}(a) shows that \itwokit is more memory
efficient when running less than (approx.) 30 Pods per Worker Node.
Note that the Kubernetes web page~\cite{large-k8} does not recommend
running more than 30 Pods per Worker Node.
Therefore, we can conclude that the memory consumption of \itwokit
behaves very well compared to Kubernetes.
In fact, we were not able to create with Kubernetes more than 42 Pods
on the same Worker Node running on a \emph{t2.xlarge} AWS Machine.
Moreover, the data reported in Table~\ref{tab:comparison}(a) does
not take into account the memory consumption of Master Nodes, which
would report a more favorable comparison to \itwokit.

Finally, there is a very active research effort targeting VM
optimization~\cite{kata,light-vm} which \itwokit can leverage in terms
of memory usage.
Unikernels~\cite{Madhavapeddy:Unikernels}, for example, are very
promising on this area. 
Also, other virtual machine optimizations have been studied in the
context of programing languages~\cite{vm-prolog}.

\subsubsection{Network Performance}
\label{sec:network-performance}  

Table~\ref{tab:comparison}(b) shows the network performance
comparison between \itwokit and different Kubernetes configurations.
The experiment uses \emph{iperf2} to measure the average network
bandwidth consumed by each Pod, where each Pod runs an \emph{iperf2}
server.
On the other hand, the \itwokit configuration runs every \emph{iperf2}
server in its own \emph{t2.large} AWS Machine.
In the table, \emph{K8-N} stands for a Kubernetes cluster with
\emph{N} Worker Nodes, where every Worker Node runs on a
\emph{t2.large} AWS Machine.
In order to be able to measure the consumed bandwidth, every
experiment runs a large amount of \emph{iperf2} clients, where each
client runs on its own VM.
These clients first send traffic to warm the load balancers up, and
then synchronize to sending traffic at the same time for 3 minutes.

% \begin{table}[t]
%   \centering
%   \begin{tabular}{|l|c|c|c|c|c|} 
%     \hline
%      Pods &   1      & 5   & 25
%     \\\hline\hline                                
%     \itwokit     & 129.86 Mbps & 128.191 Mbps  & 128.58 Mbps
%     \\\hline                                                             
%     K8-1      & 129.17 Mbps  & 25.92 Mbps   & -
%     \\\hline                                                             
%     K8-5      & 108.44 Mbps & 108.36  Mbps & 21.73 Mbps
%     \\\hline                                                             
%     K8-25      & 97.95 Mbps  & 98.11 Mbps & 97.84 Mbps
%     \\\hline                                                             
% \end{tabular}
% \vspace{5mm}
% \caption{Network comparison of \itwokit vs Kubernetes.}
% \label{tab:comparison-network}
% \end{table}

Table~\ref{tab:comparison}(b) indicates that \itwokit scales
linearly on the number of Pod replicas, as expected.
The network overhead of using an AWS Load Balancer is negligible.
Note that the limit of the virtual machine incoming traffic is 130
Mbps.
The row \emph{K8-1} in Table~\ref{tab:comparison}(b) shows that
the overhead imposed by Kubernetes when running on a single node is
not very relevant (approximately 1-2\%).
Since \emph{K8-1} runs all Pod replicas on the same machine, running
more than one Pod replica quickly hits the VM incoming bandwidth
limit.
Moreover, we were not able to successfully run 25 Pods on a single
Worker Node.
The row \emph{K8-5} shows that Kubernetes imposes an overhead of about
20\% when the Kube-Proxy needs to forward traffic between five
different Worker Nodes.
As expected, the overhead grows with the cluster size, as we can see
in the \emph{K8-25} row, which accounts for a 30\% network overhead.
Also, \emph{K8-5} shows how the traffic is dramatically affected by
the virtual machine incoming bandwidth limit when running 25 Pod
replicas.

\subsubsection{Resource Fragmentation}
\label{sec:resource-fragmentation}

The resource fragmentation suffered by \itwokit is implicitly higher
than Kubernetes, since the size of the smallest VM in AWS is 512MB.
However, this figure is competitive in production environments where
services tend to consume on the order of Gigabytes.
Also, container-based serverless architectures~\cite{faas} are a
promising option for alleviating \itwokit resource fragmentation.

On the other hand, sharing a host between different Pods imposes
performance side effects on the rest of Pods running on the same host.
Although some research has been done in this area,~\cite{quasar,
  bubble-up}, in practice IT operators reserve fix memory and CPU
resources for every Pod, introducing a similar resource fragmentation
than \itwokit.

As a final note, virtual machines running Worker Nodes tend to waste
additional resources because Worker Nodes do not run at full capacity
all the time, which introduces another level of resource
fragmentation. 
The tool \itwokit creates VMs on demand, not spawning more virtual
machines than needed, and avoiding the idle Worker Node problem
altogether.

\subsubsection{Booting Times}
\label{sec:booting-times}

The creation of a virtual machine in AWS takes about one minute, while
creating a Pod in Kubernetes takes only seconds.
Even though this difference is very relevant in local environments, it
is less relevant on production environments.
For example, it is a common practice to introduce at least a 30
seconds delay between Pod creations during a rolling update in order
for load balancers to have enough time to be updated, which induces a
comparable delay to the time required to create a virtual machine.
In summary, even though \itwokit is slower than Kubernetes in terms of
booting times, we argue that difference is not very relevant in
production environments.

\subsubsection{Security}
\label{sec:security}

There is an intrinsic security concern about running Pods from
different applications on the same Worker node~\cite{docker-security}.
Some use cases require higher level of isolation, like sandboxes for
running vulnerable or untrusted code, or multi-tenant environments in
the case of hosted services.
Container isolation uses concepts like namespaces, cgroups, seccomp
technologies, the user core linux permission model or root user
capabilities.
These mechanisms provide an additional defense on top of application
security (and they have helped to mitigate some kernel
vulnerabilities~\cite{docker-non-vulnerabilities}), but it only takes
a single kernel bug to bypass all these mechanisms and escape the
container isolation model (see~\cite{kernel-vulnerabilities} for some
vulnerabilities).

This is a strong point in favor of the isolation that \itwokit
provides compared to Kubernetes (see also~\cite{kata}).
The approach of \itwokit is to isolate Pods using secure
virtualization technology.
Note that a malicious Pod that takes control of a Worker Node could
leak information about any application running on the same Kubernetes
cluster. 
Advanced attacks can also be done by simply checking the performance
of the Pod, as it has been studied in the context of web
browsers~\cite{vila17loophole}.
Finally, \itwokit uses Linuxkit to install just the minimum set of
dependencies to run the user Pods, greatly reducing the attacking
surface of the system.
  
\section{Conclusions and Future Work}
\label{sec:conclusions}

This paper has presented \itwokit, a deployment tool that pursues the
following main goals:
\begin{inparaenum}[(a)]
\item to preserve the docker development workflow untouched;
\item to reuse popular parts of Kubernetes by accepting a simplification
of Kubernetes Manifest Files, which also eases the adoption of
\itwokit;
\item to eliminate the complexity of maintaining a Kubernetes cluster by
running a single Pod per virtual machine; and
\item to improve the security concerns of running containers in
production.
\end{inparaenum}

The \itwokit tool eliminates some of Kubernetes complexity for
production environments by running on public cloud vendors.
The results in Section~\ref{sec:exp} suggest that the memory
consumption of \itwokit is comparable to Kubernetes, that the network
performance of \itwokit is better than the one of Kubernetes, and that
the price of having slower booting times and larger resource
fragmentation seems acceptable in production environments.
%
% We still think Kubernetes is particularly valuable in private
% data-centers, where user would otherwise have to manage the complexity
% of other solutions such as Openstack or VMware technology.
  
The tool \itwokit is a deployment tool that tries to exploit synergies
between the world of containers (widely used for development) and the
world of virtual machines (widely used for production).
Our tool creates a feedback loop where developers are able to improve
their local workflows using containers, and operators transform those
containers into production-ready minimal virtual machines.
Current and future research includes trying to optimize further the
generation of minimal virtual machines by reducing the size of our
base Linuxkit distribution, and by using Unikernels.
Another research line is to integrate \itwokit with server-less
architectures~\cite{faas} provided by cloud vendors.

Finally, \itwokit not only reduces the complexity of the deployment
system by using cloud vendor technology not managed by the end user,
but \itwokit also brings a significant security improvement for two
reasons.
First, malicious Pods are isolated by secured and proven hypervisor
technology.
Second, the use of specialized Linuxkit distributions reduces the
attacking surface of the resulting system.
The tool \itwokit can be improved even further to limit the set of
capabilities needed by the user Pods.
For example, the \emph{nginx} container of the Linuxkit template shown
in Fig.~\ref{fig:nginx-linuxkit} could also limit its capabilities
to \texttt{CAP\_NET\_BIND\_SERVICE}, \texttt{CAP\_CHOWN},
\texttt{CAP\_SETUID}, \texttt{CAP\_SETGID} and
\texttt{CAP\_DAC\_OVERRIDE}.

\Urlmuskip=0mu plus 1mu\relax
\bibliographystyle{IEEEtran}
\bibliography{main}

\end{document}